\documentclass[aps,prl,twocolumn,showpacs,superscriptaddress,groupedaddress]{revtex4-1}  
\usepackage{graphicx}  
\usepackage{dcolumn}   
\usepackage{bm}        
\usepackage{amssymb}   
\usepackage{amsmath}
\usepackage{xspace}
\usepackage[caption = false]{subfig}
\usepackage{ulem} 

\usepackage{times}
\usepackage{color}
\usepackage{ulem}

\newcommand{\be}{\begin{equation}}
\newcommand{\ee}{\end{equation}}

\begin{document} 
\title{Super-Penrose collisions are inefficient - a Comment on: Black hole fireworks: ultra-high-energy debris from super-Penrose collisions  }


\author
{Elly Leiderschneider and Tsvi Piran }
\affiliation{ Racah Institute of Physics, The Hebrew University of Jerusalem, Jerusalem 91904, Israel }
\date{\today}

\begin{abstract}
{\bf Abstract:}  
In a paper posted on the arXiv a few weeks ago Berti,  Brito and  Cardoso   \cite{Berti+14}
suggest that ultra-high-energy particles can emerge from collisions in a black hole's ergosphere. 
This can happen if the process involves a particle on an outgoing trajectory very close to the black hole. Clearly such a particle cannot emerge from the black hole.  It is argued \cite{Berti+14} that this particle can arise in another collision. Thus the process involves two collisions: one in which an outgoing particle is produced extremely close to the horizon, and a second one in which energy is gained. The real efficiency of this process should take into account, therefore,  the energy needed to produce
the first particle. We show here that  while this process is kinematically possible, it requires a deposition of energy that is divergently large compared with the energy of the escaping particle. Thus, in contradiction to claims of infinitely high efficiencies, the efficiency of the combined process is in fact extremely small, approaching zero for very high output energies. Even under more general conditions than those considered in \cite{Berti+14} the total energy gain never diverges, and is larger only by a factor of a few than the energy gain of the original collisional Penrose process 
that takes place between two infalling particles \cite{Piran+75,PiranShaham77,Bejger+12}.
\end{abstract}
\maketitle

\section{Introduction}
\label{sec:intro}

\begin{figure*}[]
\includegraphics[width=43mm]{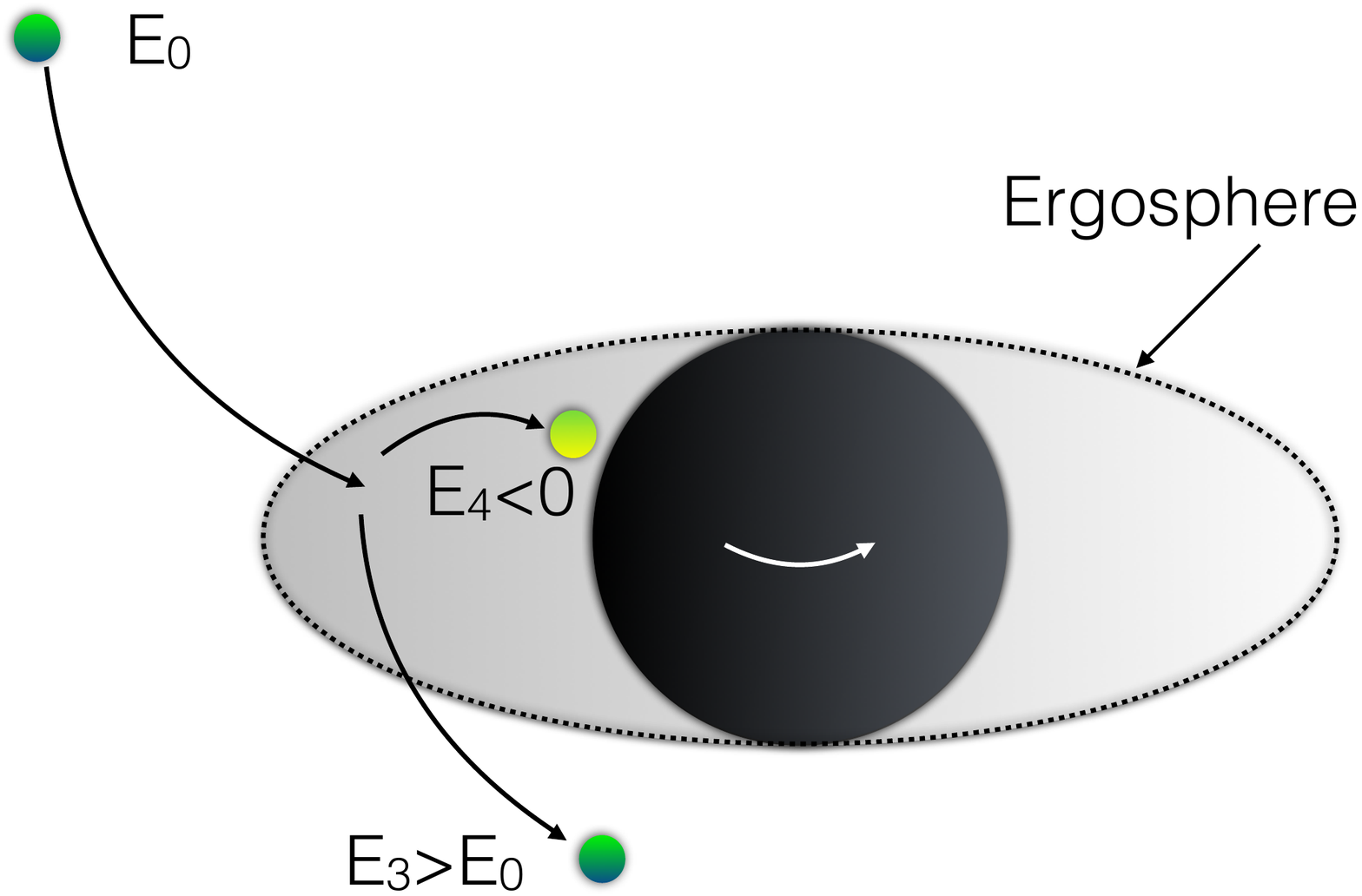}
\includegraphics[width=43mm]{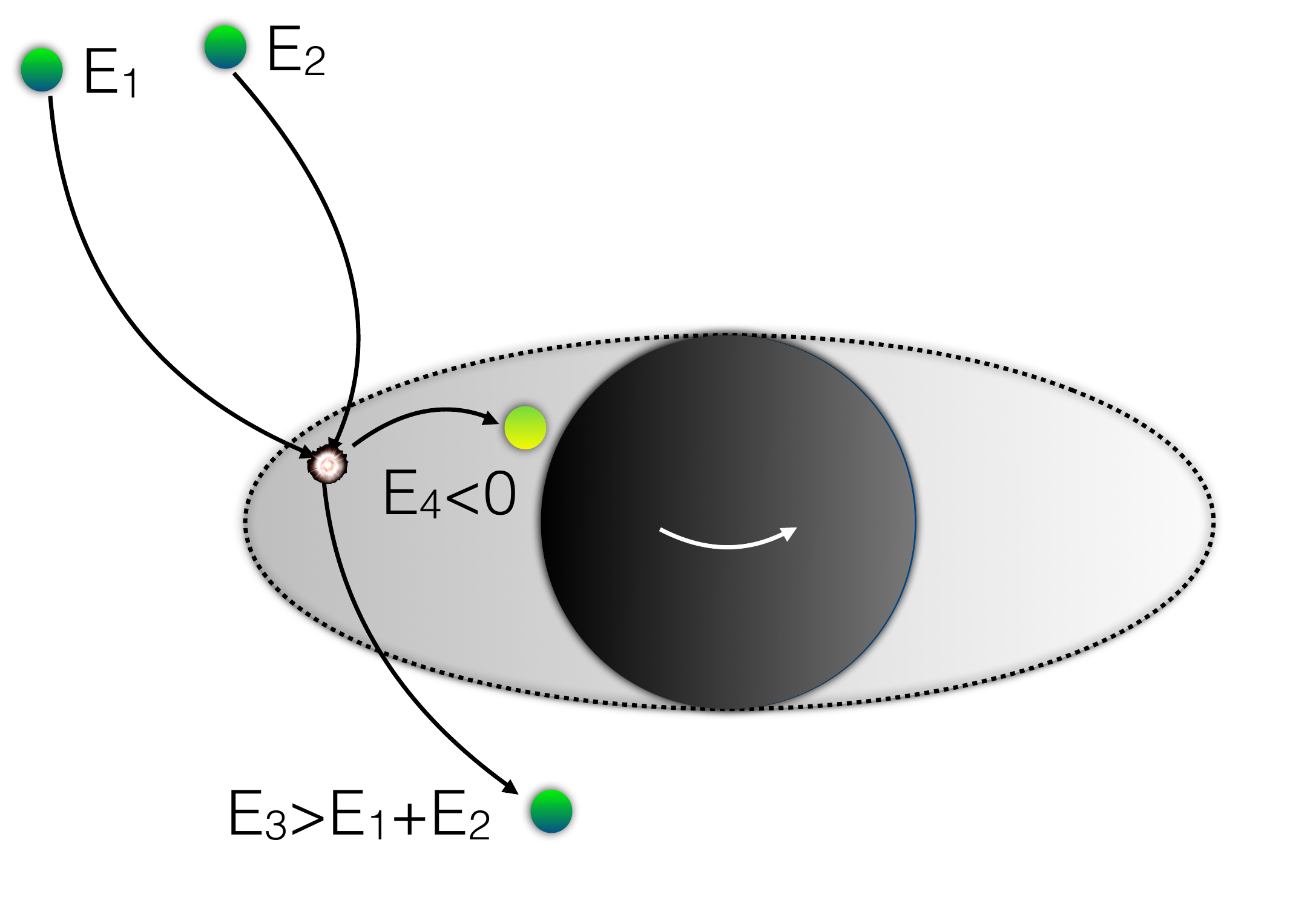}
\includegraphics[width=43mm]{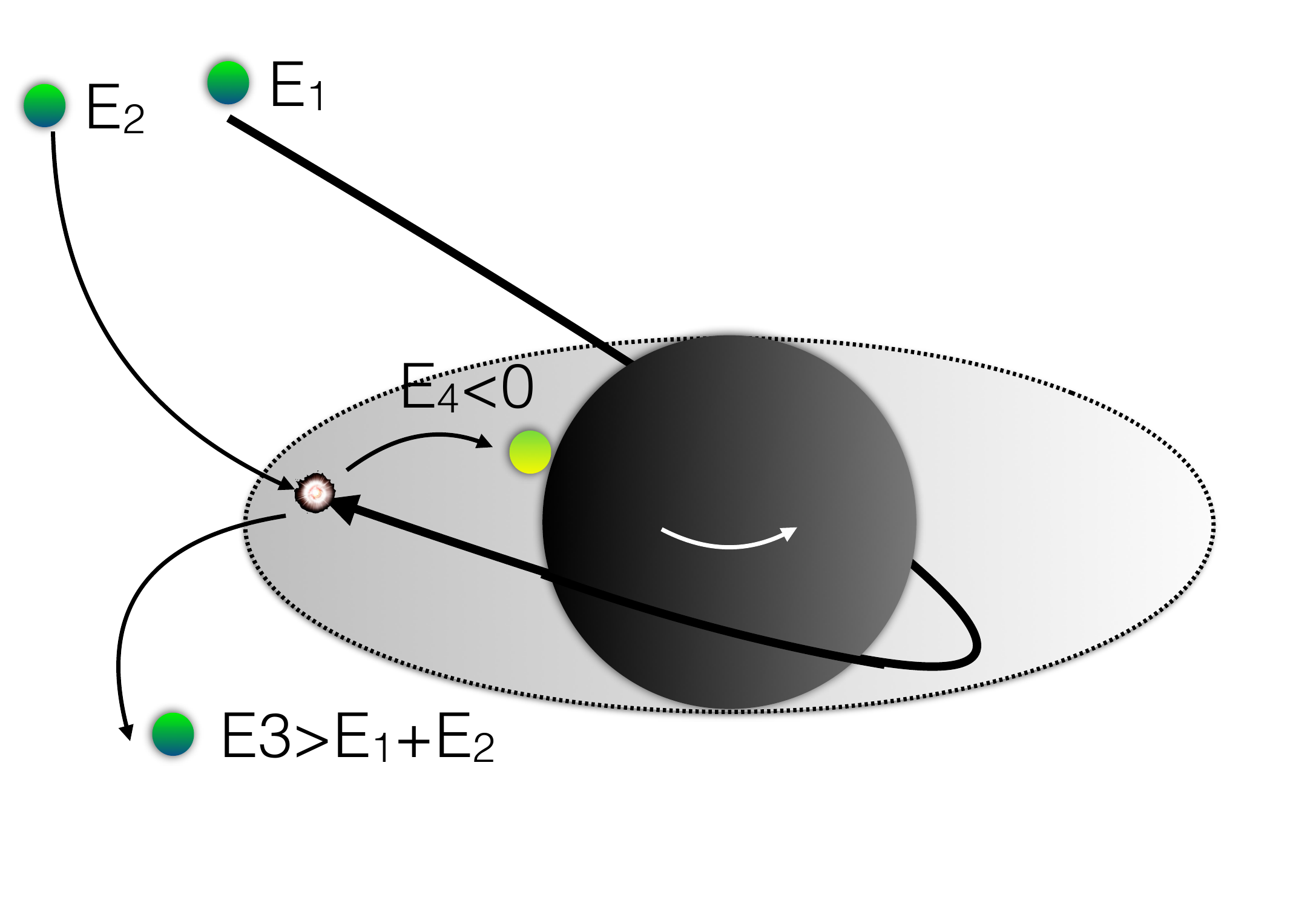}
\includegraphics[width=43mm]{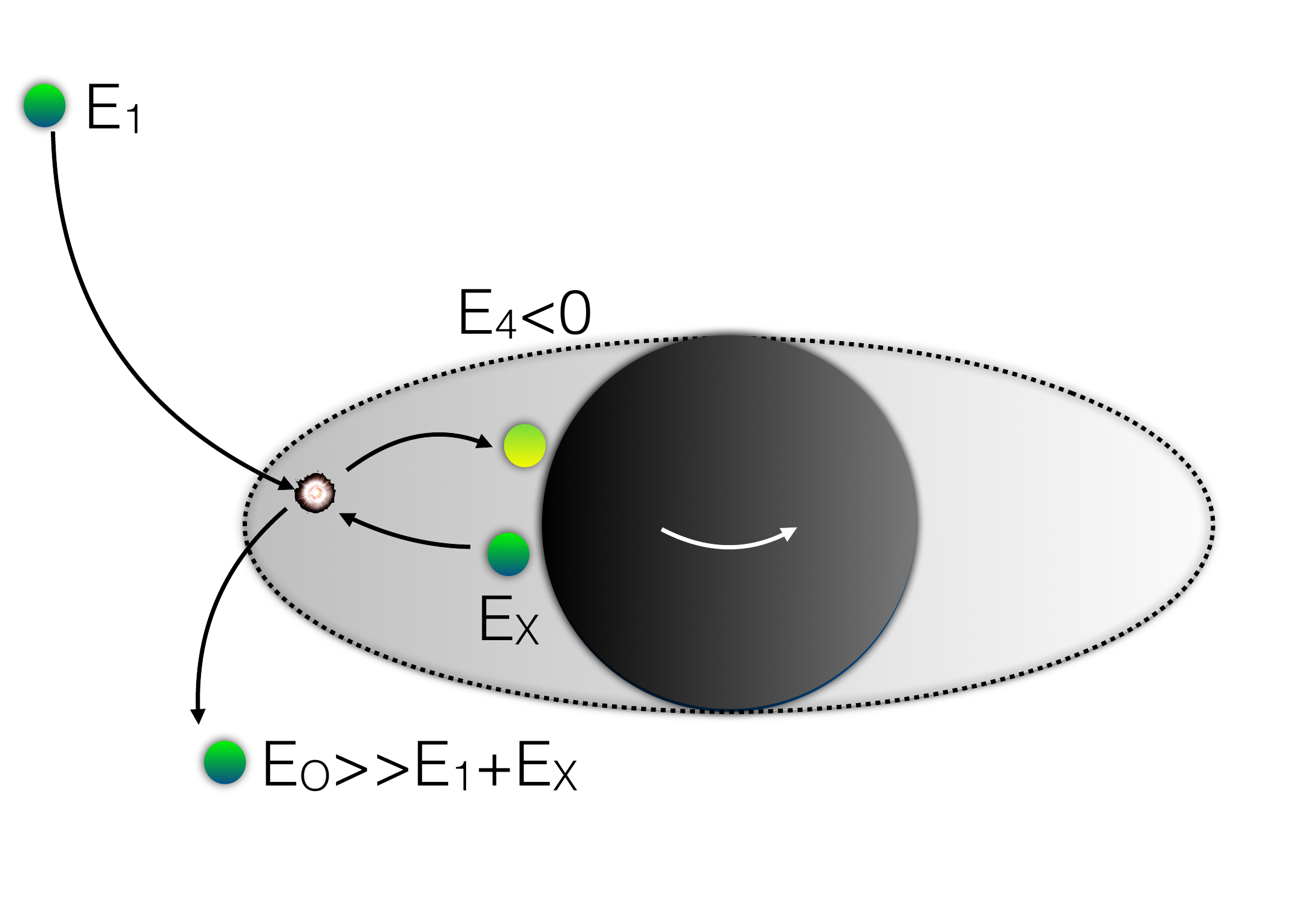}
\caption{ Left: The original Penrose process - a particle disintegrates inside the ergosphere. One of the resulting particles possesses negative energy, and falls into the black hole. The other one escapes to infinity with more energy than the energy of the initial particle \cite{Penrose69}. Center left: A collisional Penrose process. Two infalling particles collide in the ergosphere \cite{Piran+75}.
Center right: One of the infalling particles turns around the black hole and collides with a second infalling particle while it is outgoing \cite{Schnittman14}. Right: An infalling particle $_1$ collides with an outgoing particle $_X$ emerging from the black hole \cite{Berti+14}. }
\label{fig:penrose}
\end{figure*}

The original Penrose process \cite{Penrose69} (see Fig. \ref{fig:penrose}) involved particle disintegration in a Kerr black hole ergosphere. One of the resulting particles falls into the black hole on a negative orbit, while the other escapes to infinity with energy larger than the original energy of the infalling particle. 
The energy gain arises, of course, from the rotational energy of the Kerr black hole that absorbs the negative energy particle.
Shortly after this process was proposed, it was shown \cite{Bardeen+72,Wald74, KovetzPiran75} that a significant fraction of the infalling particle's rest mass must be converted to energy in its rest frame, in order that the positive energy particle escapes to infinity. The energy gain from the black hole is not very significant in this case compared with the energy conversion in the particle's rest frame, making this process ``somewhat uninteresting" from an astrophysical or technological\footnote{There was a suggestion that an advanced civilization would power itself by dumping its garbage on a Kerr black hole and extracting, via the Penrose process, the black hole's rotational energy}  point of view.

Subsequently Piran, Shaham and Katz \cite{Piran+75} have shown that a collisional Penrose process (see Fig. \ref{fig:penrose}) could work. In a collision between two infalling particles that takes place within the ergosphere, the energy in the CM frame can be large (in fact, it can be infinitely large under suitable conditions), and as such, it is possible that a Penrose process could take place with positive energy gain for a particle that escapes to infinity. Detailed analysis \cite{PiranShaham77}, however, has shown that even though the energy in the CM can grow infinitely large, when imposing the condition that one of the particles escapes to infinity, the energy gain would still be modest (of order unity).


Recent interest in this idea arose when Banados et al., {Banados+09} rediscovered that a collision near a Kerr black hole's horizon can have infinitely large energy in the CM frame. It was suggested that this infinite energy could be utilized to accelerate particles to extremely large energies. However, detailed studies \cite{Bejger+12} refined some of the earlier analysis of Penrose collisions.  These studies have shown, once more, that  the requirement that the particle produced in the collision 
escapes to infinity imposes stringent conditions on its trajectory, and limits its energy to, at most, a few times the energy of the infalling particles. This result is intuitive: when two infalling particles collide near the horizon, their CM has an enormous negative radial momentum. Hence, most of the particles produced in such  collisions will fall into the black hole, and only a small fraction will escape. 

These last estimates assumed, naturally, that the colliding particles (that are deep within the ergosphere and, for high efficiency, infinitely close to the black hole's horizon) are on in-going trajectories. This natural assumption was revised by Schnittmann \cite{Schnittman14}, who pointed out that a particle with sufficient angular momentum can infall from infinity and bounce to an outgoing orbit, even infinitely close to the horizon. Now we can have a collision between two particles, originally infalling from infinity, one of which is outgoing 
(see Fig. \ref{fig:penrose}). Clearly the radial momentum is larger in this case, and the probability of the resulting particle to escape is larger. Correspondingly, a larger energy gain by a factor of a few (compared to the case of two infalling particles) is possible \cite{Schnittman14}.

More recently, Berti et al. \cite{Berti+14} considered even more radical initial conditions. They also consider a collision between an outgoing and infalling particle- however, this outgoing particle, which we denote here with a subscript $_X$,  is not one that has fallen from infinity and turned around the black hole. The angular momentum of this particle is smaller than the critical one needed for a particle to bounce. Instead, this particle is emerging outwards from the black hole (see Fig. \ref{fig:penrose}).
A collision of such a particle with an infalling particle can result in an outgoing particle whose energy is unbounded. In the following, we will denote this particle by the subscript $_O$(for "outgoing"), indicating that this particle is an escaping one. Berti et al \cite{Berti+14} dub such collisions "super-Penrose," and suggest that ultra-high-energy debris can arise from them.

However, one should ask - can such an outgoing particle occur naturally? Berti et al., \cite{Berti+14} address this question, and show that a collision between two particles, infalling from rest at infinity, can result in an $_X$ type particle. This comes at a hefty price, however- one of the infalling particles must be extremely energetic. We show here that the energy needed to produce the essential intermediate particle $_X$  is significantly larger than the energy of the escaping final particle $_O$. Thus, from the point of view of an observer at infinity, the super-Penrose collision results in a net (extremely large) energy loss- the black gains energy, rather than giving its energy away.

\section{Efficiency of Penrose Collisions}

For simplicity, we consider a maximal Kerr black hole ($a/m=1$, for which the Penrose process is most
effective), and collisions that take place in the equatorial plane. 
We define $m_i$, $E_i$ and  $L_i$ as the mass, energy and angular momentum of particle $i$, as measured by an observer at rest at infinity. 
We denote the impact parameter of a particle as $b_i \equiv L_i/E_i$, and note that particles with $b_i > b_{crit} = 2$ (and a  positive $L_i$) are deflected and turn around the black hole, unable to reach $r=1$. With $b=2$ a particle is  deflected at $r=1$ (but, because of the singular nature of the radial coordinate at the horizon for a=1, this  deflection point is still outside the horizon).

\begin{figure*}[]
\vspace{-1.0cm}
\includegraphics[width=60mm]{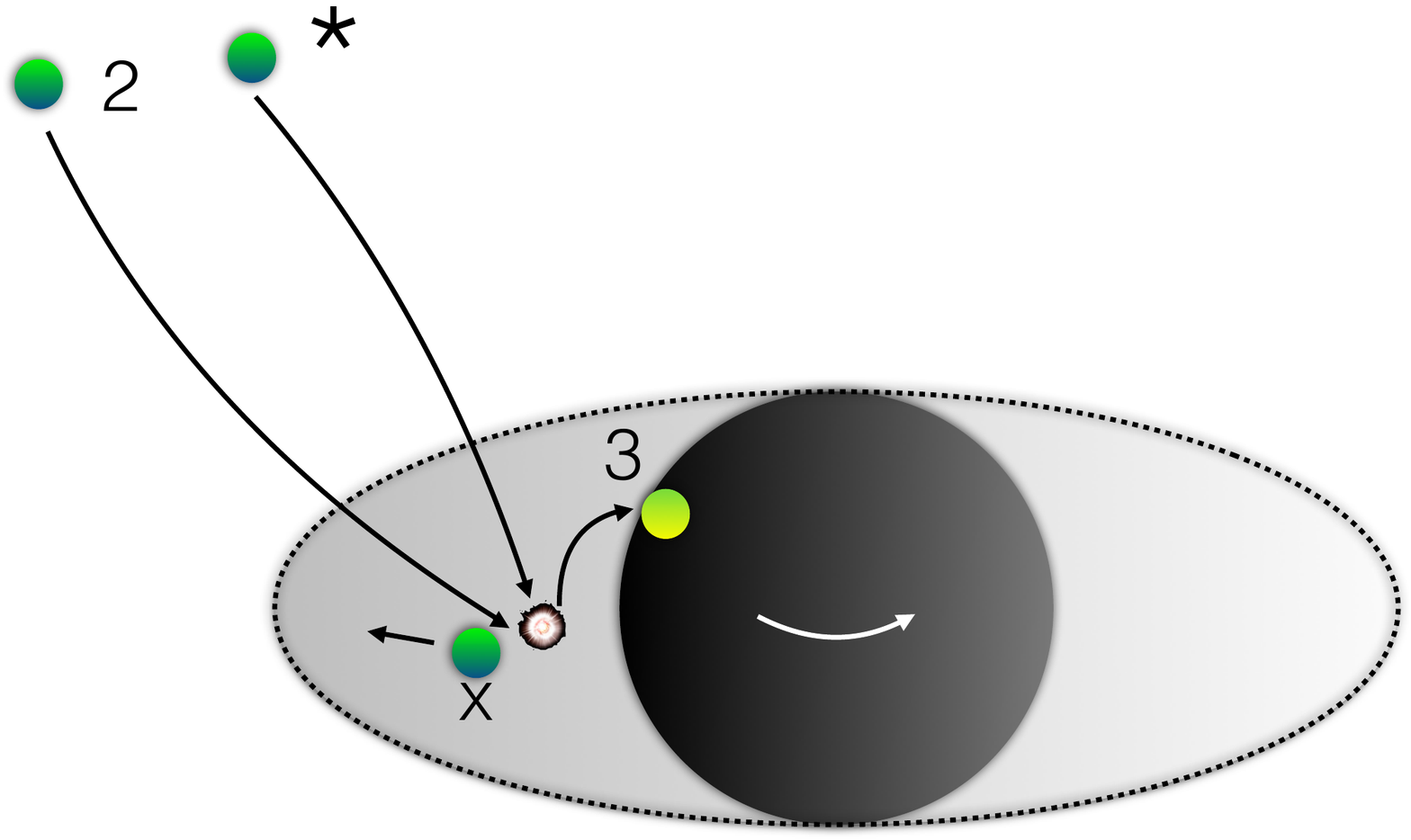}
\hspace{-1cm}\includegraphics[width=60mm]{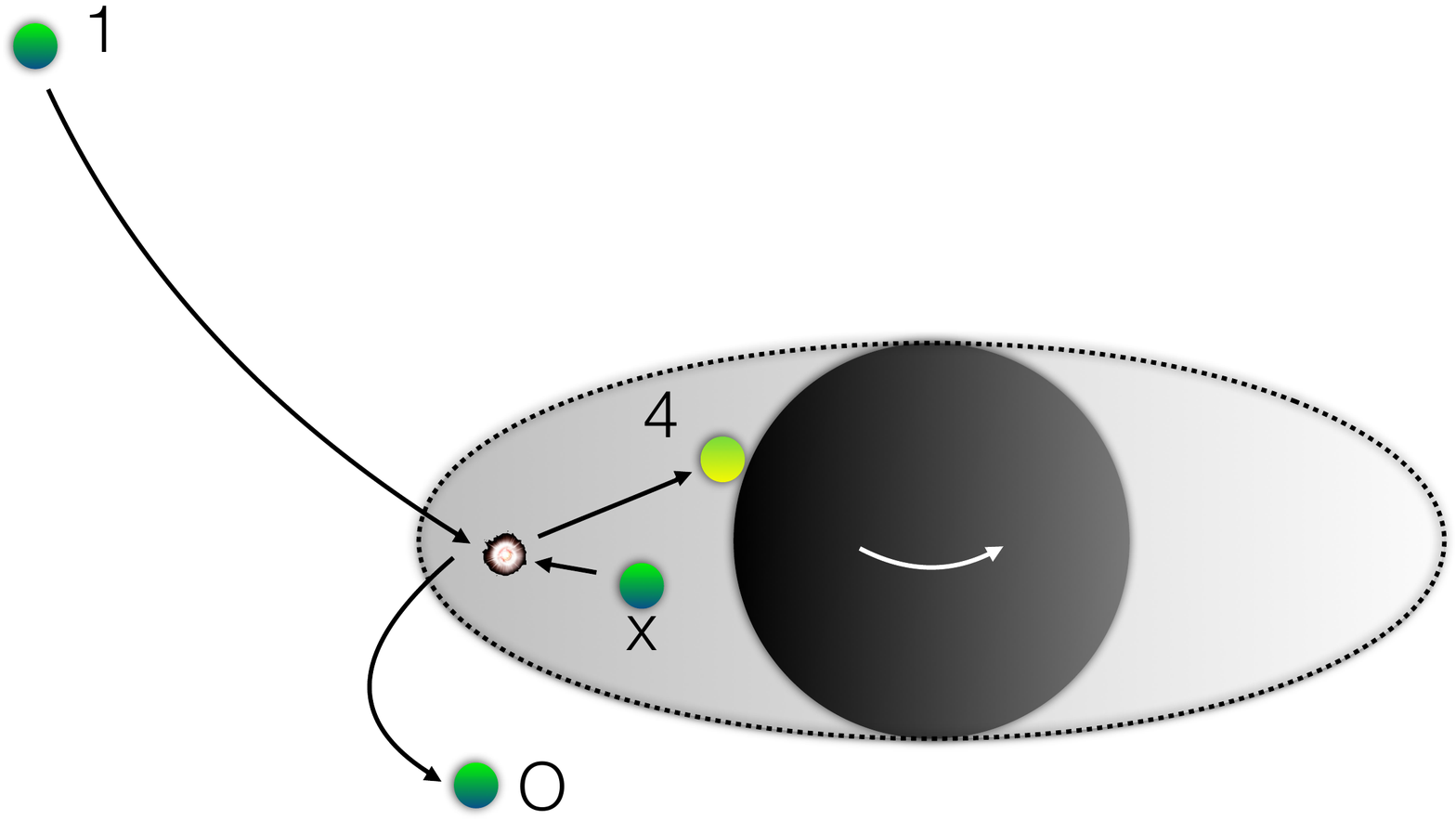}
\hspace{-1cm}\includegraphics[width=60mm]{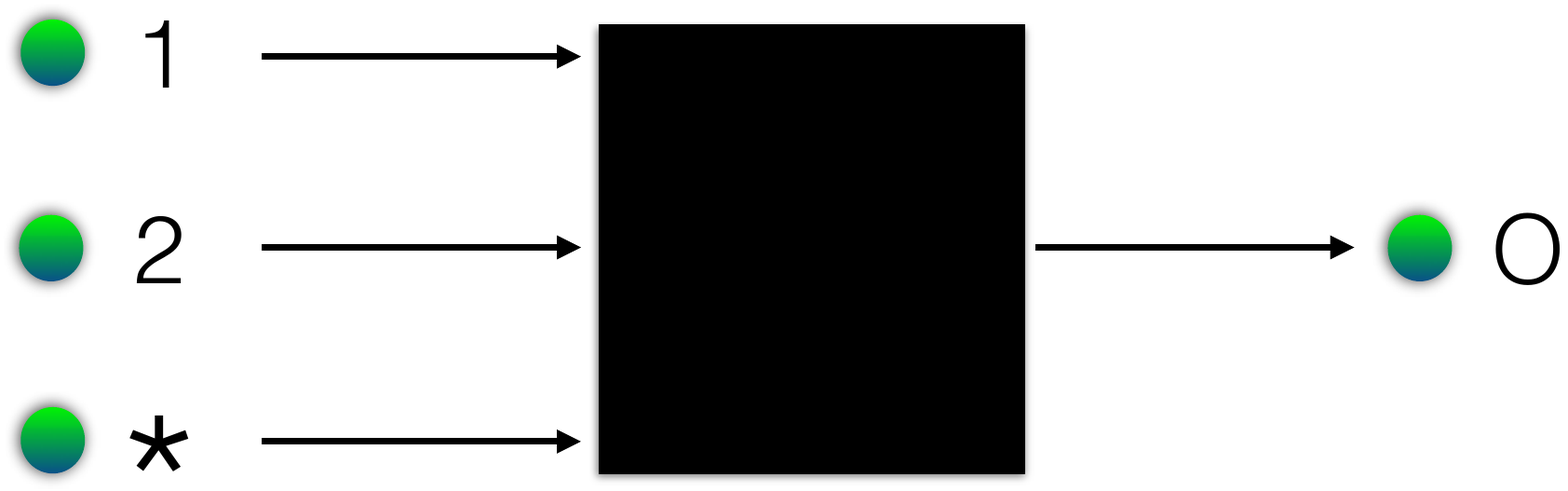}

\caption{The two successive collisions involved in the "super-Penrose" process.  Left: The first collision between two infalling particles (one of which, $_*$,  is very massive) produces an outgoing particle $_X$ and an infalling particle. Center: The second collision between the outgoing particle $_X$ and another infalling particle, produces an energetic outgoing particle $_O$ and an infalling particle. Note that the figure is schematic and both collisions have to take place very close to the black hole's horizon.  Right: Schematically for an observer at infinity, the two collisions should  be considered as a single process in which three particles $_1$, $_2$ and $_*$ fall in and a fourth particle $_O$ gets out. }
\label{fig:penrose3}
\end{figure*}
We consider two  successive collisions (see Fig. \ref{fig:penrose3}). 
{The first collision is between particle $_2$ and particle $_*$ that fall from infinity, collide, and produce an outgoing particle denoted $_X$ as well as another particle (marked $_3$ in Fig. \ref{fig:penrose3}) that falls into the black hole. The   second collision is  the  super-Penrose process  discussed in \cite{Berti+14}. In this collision  an infalling particle, denoted $_1$, and the outgoing particle  $_X$, that was produced in the first collision, collide and annihilate, producing two photons:  one, denoted $_O$ that  escapes to infinity.  and another one that falls into the black hole (marked $_4$ in Fig. \ref{fig:penrose3}). }

We begin by examining the second collision, the super-Penrose one. Particle $_X$ is outgoing, and its impact parameter $L/E$ is less than the critical one.  To normalize the calculations, we choose $E_1=E_X=1$ and $m_1=m_X=1$. The outgoing particle $_X$ has an impact parameter $b_X<2$. 
Fig.  3
depicts the maximal efficiency of this collision, defined as: 
\begin{eqnarray}
&&\eta_2(r ,b_1,b_0)\equiv \frac{{\rm Max}[E_O(r ,b_1,b_x)]} {E_1+E_X} = \\
\nonumber
&&\frac{( 2 - {b_1})\, ( 2 - {b_X})}
{ 2 ( 4 - {b_1} -  {b_X} ) + \sqrt{3}\, ({b_1} - {b_X})}\cdot \frac{1}{(r - 1)} +O (1),
\label{Eq:eff1}
\end{eqnarray}
where  ${\rm Max}[E_O(r ,b_1,b_0)]$ is the maximal possible energy of particle $_O$ that escapes to infinity. 
This efficiency, $\eta_2$, diverges like $1/(r-1)$ as the collision point approaches the horizon ($r \rightarrow 1 $).

As mentioned earlier, the origin of particle $_X$  is non-trivial. $E_X=M_X$ suggests a massive particle falling from rest at infinity, but an infalling particle with $b<2$ will  fall into the black hole. Berti et al. \cite{Berti+14} propose that such a particle can be produced in a near-horizon collision of two other massive particles, falling from rest at infinity. We denote these particles as $_2$ and $_*$. 
The collisions results in two particles: $_X$, and a particle denoted $_3$ (which ultimately falls into the black hole). 

\begin{figure*}[]
\includegraphics[width=100mm]{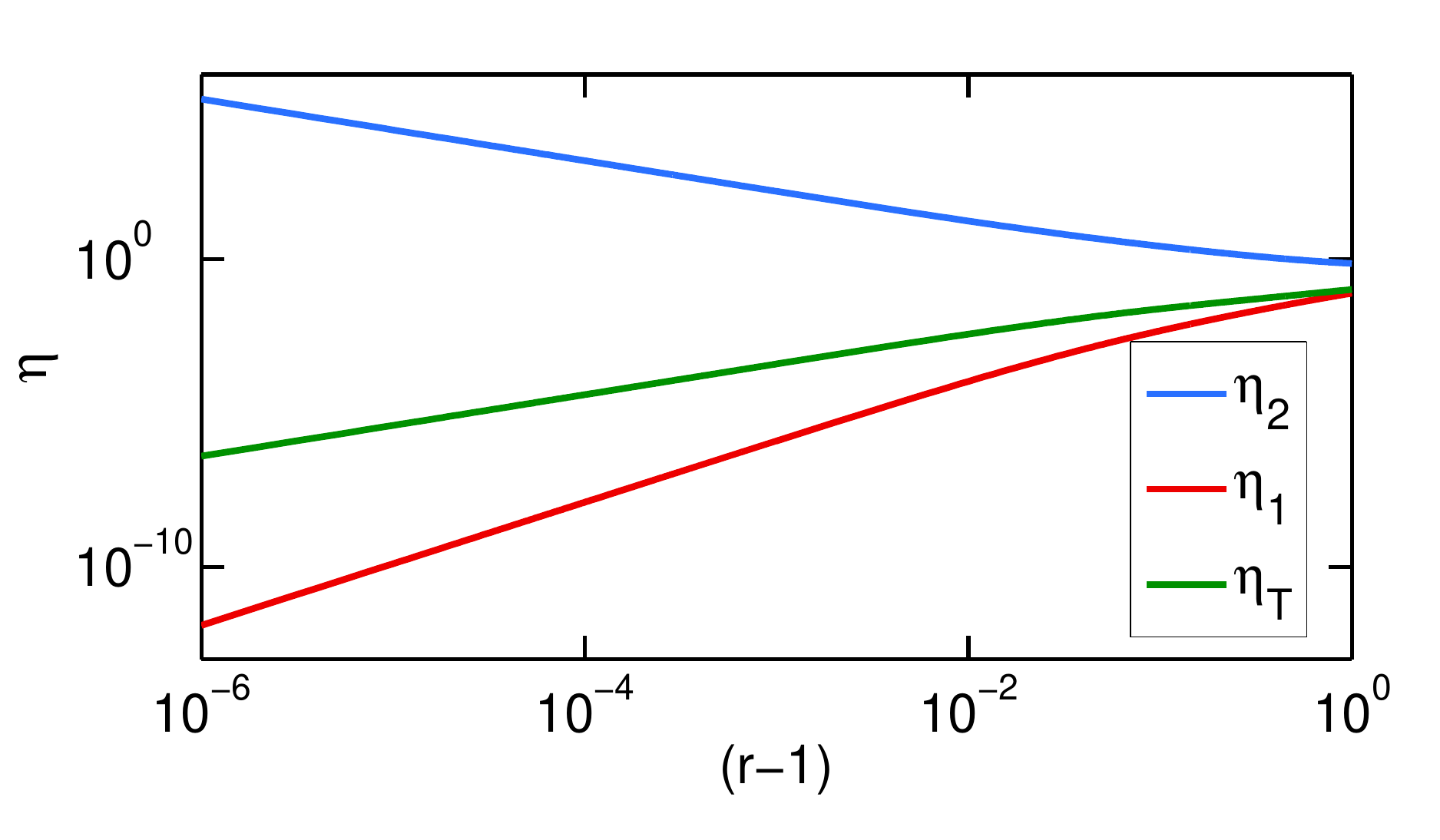}
\vspace{-.50cm}
\label{fig:eff1}
\caption{{ The efficiency of the two collisions, $\eta_1$ and $\eta_2$ and the total efficiency $\eta_T$. 
In this example we have chosen $b_1=1.9$ and $b_2=1$. These values determine the specific values shown here but not the generic behavior.
The second collision, that is very efficient, is the one discussed by \cite{Berti+14}. This efficiency diverges as the collision point approaches the horizon.  The  efficiency of the first collision, that produces the outgoing particle $_X$,   is $\eta_1$ and $\eta_T$ is the total efficiency. This total efficiency, $\eta_T$ vanishes as the collision point approaches the horizon.  }}
\end{figure*}

Following \cite{Berti+14}, we choose $_2$ and $_*$ to radially fall from rest  at infinity with no angular momentum: $L_2=L_*=0$.  But while $m_2=1$,  $ m_*$ is unspecified. 
We calculate the minimal mass,  $m_*$,  required  to produce the outgoing sub-critical particle $_X$ with $E_X=m_X=1$, $b_X<2$ and $p^r_X>0$ (together with an infalling particle $_3$). To do so, we solve,  bearing in mind conservation of energy and angular momentum, the radial momentum equation:
\begin{equation}
p^r_2 + p^r_* = p^r_X + p^r_3
\end{equation} 
where
\begin{equation} 
p^r_i = \sqrt{\frac{E_i^2-m_i^2}{2} + \frac{m_i^2}{r} - \frac{L_i^2-(E_i^2-m_i^2)}{2 r^2} + \frac{(L_i-E_i)^2}{r^3}} ,
\end{equation} 
where we have set here $M_{_{BH}}=a=1$.
We find that: 
\begin{equation}
m_*(r,\epsilon) \ge \frac{8 (2-L_X)}{(r-1)^2}+O ((r-1)^{-1}) ,
\end{equation}  
where the equality is satisfied in the limiting case of $m_3=0$.   
{The corresponding efficiency of this collision, $\eta_1$, is:}
\begin{equation}
\eta_1 \equiv \frac{E_X}{m_2+m_s} = \frac{1}{1+m_*} \le \frac{(r-1)^2}{8 (2-L_X)} + O(r-1) \ . 
\label{eq:eff1}
\end{equation}


The crux of the problem is that for an observer at infinity, the energy loss involves the rest mass energies of  particles infalling from infinity:
$m_1$, $m_2$ and $m_*$, while  the energy gain is $E_O$. Thus, the real efficiency of the whole process is the product of two efficiencies -  The efficiency of the first collision, $ \eta_1$, and the efficiency of the second collision, $\eta_2$. All together we have: 
\begin{eqnarray}
&&\eta_{T} \equiv \frac{{\rm Max}[E_O(r ,b_1,b_X)]} {E_1+E_2+E_*} = \frac{{\rm Max}[E_O(r ,b_1,b_X)]} {2+m_*}  \\
\nonumber 
&&=\frac{2- b_1}{4[ 2 (4-  {b_X} - {b_1} )+ \sqrt{3}\, \left({b_X} - {b_1}\right) ]}\cdot (r-1) +O (r-1)^2
\   .
\end{eqnarray}
This $\eta_T$ should replace $\eta_2$, given by  \cite{Berti+14}
as the real efficiency of the process.
While $\eta_2$ diverges as $\rightarrow 1$, $\eta_T$ becomes diminishingly small, and vanishes as $(r-1)$.

\section{Conclusions}
To summarize we have shown that while the maximal energy of the escaping particle $_O$  diverges, infinitely large energy has to be invested in order to produce it.  The ratio of the energy output compared with the energy invested is vanishingly small. Namely  this process is extremely inefficient.  {The maximal efficiency in this combined process $\eta_{T_{max}} = 1-\sqrt{(5+\sqrt{5})/10}=0.1493$  is less than unity. This happens when $b_1 \rightarrow 2$ and $b_2 \rightarrow -2(1+\sqrt{ 2})$ and the collision takes place on the boundary of the ergosphere. }

It should be stressed that this limit to the efficiency holds for the specific configuration discussed by \cite{Berti+14} in which $m_1=m_2=1$ and $b_1 = b_* = 0$ while $m_*$ and   $b_2$ are arbitrary. 
One can generalize the process and consider arbitrary impact parameters, $b_1$, $b_2$  and $b_*$   for all three particles (keeping however $m_1=m_2=1$). If $b_i\ge 2$ one can also consider the possibility that this particle  has turned around the black hole and it is outgoing at the point of the collision.  This general case allows for larger efficiencies and one can see that the overall efficiency, $\eta_T$ in this case is slightly larger but comparable to the maximal efficiency obtained by \cite{Schnittman14}. As expected, this indeed happens when both collisions are close to the horizon and one of the particles has turned around the black hole and is moving outwards. However, given that this ``black box" of three incoming particles and two internal collisions is much more complicated and requires much higher fine tuning than the single collisions considered by \cite{Schnittman14}, this gain in efficiency is unlikely to be of any physical relevance.  

The basic difficulty in obtaining an efficient collisional Penrose process is that for a realistic collision that takes place between two infalling particles, the overall momentum of the CM system points towards the black hole. The condition that a particle produced in such a collision escapes from the black hole is very limiting, and correspondingly the energy gain is not large. Clearly, if one of the colliding particles is on an outgoing trajectory, the overall ingoing momentum of the system is reduced and a particle emerging from this collision can escape more easily. Hence, higher energy gains are possible. The configuration is still limited if one allows that the outgoing particle is an infalling one that has turned around the black hole \cite{Schnittman14}. In this case, the collision must be close to the deflection point of this particle, and hence its radial momentum must be small. A much higher energy gain is possible if one invokes an arbitrary outgoing particle near the black hole. However, such a particle cannot emerge from the black hole.  It was suggested \cite{Berti+14} to produce this outgoing  particle by yet another collision. We have shown, here, that the production of such an outgoing particle is not a trivial issue. It involves enormous energy, and the final efficiency of the combined process (the production of the outgoing particle and the subsequent Penrose collision) is diminishingly small in the case considered in \cite{Berti+14}. At  best, under more general conditions, this combined process can lead only to a modest energy gain, which is comparable to the total energy deposited at infinity.  Even disregarding the fine tuning required, these considerations suggest that it is quite unlikely that this such collisions take place in realistic astrophysical configurations.

\begin{acknowledgments}
We thank M. Abramowicz and M. Bejger for  fruitful discussions.
This research was supported by an ERC advanced grant (GRBs) and by the  I-CORE 
Program of the Planning and Budgeting Committee and The Israel Science
Foundation (grant No 1829/12). 

\end{acknowledgments}

\def\apj{Astrophys.\ J.}
\def\nat{Nature}
\def\apjl{Astrophys.\ J. Lett.}
\def\apj{Astrophys.\ J.}
\def\aap{Astron.\ Astrophys.}
\def\prd{Phys. Rev. D}
\def\physrep{Phys.\ Rep.}
\def\mnras{Month. Not. RAS }
\def\araa{Annual Rev. Astron. \& Astrophys.}
\def\aapr{Astron. \& Astrophys. Rev.}
\def\aj{Astronom. J.}
\def\jcap{JCAP}


\begin{thebibliography}{10}%
\makeatletter
\providecommand \@ifxundefined [1]{%
 \@ifx{#1\undefined}
}%
\providecommand \@ifnum [1]{%
 \ifnum #1\expandafter \@firstoftwo
 \else \expandafter \@secondoftwo
 \fi
}%
\providecommand \@ifx [1]{%
 \ifx #1\expandafter \@firstoftwo
 \else \expandafter \@secondoftwo
 \fi
}%
\providecommand \natexlab [1]{#1}%
\providecommand \enquote  [1]{``#1''}%
\providecommand \bibnamefont  [1]{#1}%
\providecommand \bibfnamefont [1]{#1}%
\providecommand \citenamefont [1]{#1}%
\providecommand \href@noop [0]{\@secondoftwo}%
\providecommand \href [0]{\begingroup \@sanitize@url \@href}%
\providecommand \@href[1]{\@@startlink{#1}\@@href}%
\providecommand \@@href[1]{\endgroup#1\@@endlink}%
\providecommand \@sanitize@url [0]{\catcode `\\12\catcode `\$12\catcode
  `\&12\catcode `\#12\catcode `\^12\catcode `\_12\catcode `\%12\relax}%
\providecommand \@@startlink[1]{}%
\providecommand \@@endlink[0]{}%
\providecommand \url  [0]{\begingroup\@sanitize@url \@url }%
\providecommand \@url [1]{\endgroup\@href {#1}{\urlprefix }}%
\providecommand \urlprefix  [0]{URL }%
\providecommand \Eprint [0]{\href }%
\providecommand \doibase [0]{http://dx.doi.org/}%
\providecommand \selectlanguage [0]{\@gobble}%
\providecommand \bibinfo  [0]{\@secondoftwo}%
\providecommand \bibfield  [0]{\@secondoftwo}%
\providecommand \translation [1]{[#1]}%
\providecommand \BibitemOpen [0]{}%
\providecommand \bibitemStop [0]{}%
\providecommand \bibitemNoStop [0]{.\EOS\space}%
\providecommand \EOS [0]{\spacefactor3000\relax}%
\providecommand \BibitemShut  [1]{\csname bibitem#1\endcsname}%
\let\auto@bib@innerbib\@empty
\bibitem [{\citenamefont {{Berti}}\ \emph {et~al.}(2014)\citenamefont
  {{Berti}}, \citenamefont {{Brito}},\ and\ \citenamefont
  {{Cardoso}}}]{Berti+14}%
  \BibitemOpen
  \bibfield  {author} {\bibinfo {author} {\bibfnamefont {E.}~\bibnamefont
  {{Berti}}}, \bibinfo {author} {\bibfnamefont {R.}~\bibnamefont {{Brito}}}, \
  and\ \bibinfo {author} {\bibfnamefont {V.}~\bibnamefont {{Cardoso}}},\
  }\href@noop {} {\bibfield  {journal} {\bibinfo  {journal} {ArXiv e-prints}\ }
  (\bibinfo {year} {2014})},\ \Eprint {http://arxiv.org/abs/1410.8534}
  {arXiv:1410.8534 [gr-qc]} \BibitemShut {NoStop}%
\bibitem [{\citenamefont {{Piran}}\ \emph {et~al.}(1975)\citenamefont
  {{Piran}}, \citenamefont {{Shaham}},\ and\ \citenamefont
  {{Katz}}}]{Piran+75}%
  \BibitemOpen
  \bibfield  {author} {\bibinfo {author} {\bibfnamefont {T.}~\bibnamefont
  {{Piran}}}, \bibinfo {author} {\bibfnamefont {J.}~\bibnamefont {{Shaham}}}, \
  and\ \bibinfo {author} {\bibfnamefont {J.}~\bibnamefont {{Katz}}},\ }\href
  {\doibase 10.1086/181755} {\bibfield  {journal} {\bibinfo  {journal} {\apjl}\
  }\textbf {\bibinfo {volume} {196}},\ \bibinfo {pages} {L107} (\bibinfo {year}
  {1975})}\BibitemShut {NoStop}%
\bibitem [{\citenamefont {{Piran}}\ and\ \citenamefont
  {{Shaham}}(1977)}]{PiranShaham77}%
  \BibitemOpen
  \bibfield  {author} {\bibinfo {author} {\bibfnamefont {T.}~\bibnamefont
  {{Piran}}}\ and\ \bibinfo {author} {\bibfnamefont {J.}~\bibnamefont
  {{Shaham}}},\ }\href {\doibase 10.1103/PhysRevD.16.1615} {\bibfield
  {journal} {\bibinfo  {journal} {Phys. Rev. D.}\ }\textbf {\bibinfo {volume}
  {16}},\ \bibinfo {pages} {1615} (\bibinfo {year} {1977})}\BibitemShut
  {NoStop}%
\bibitem [{\citenamefont {{Bejger}}\ \emph {et~al.}(2012)\citenamefont
  {{Bejger}}, \citenamefont {{Piran}}, \citenamefont {{Abramowicz}},\ and\
  \citenamefont {{H{\aa}kanson}}}]{Bejger+12}%
  \BibitemOpen
  \bibfield  {author} {\bibinfo {author} {\bibfnamefont {M.}~\bibnamefont
  {{Bejger}}}, \bibinfo {author} {\bibfnamefont {T.}~\bibnamefont {{Piran}}},
  \bibinfo {author} {\bibfnamefont {M.}~\bibnamefont {{Abramowicz}}}, \ and\
  \bibinfo {author} {\bibfnamefont {F.}~\bibnamefont {{H{\aa}kanson}}},\ }\href
  {\doibase 10.1103/PhysRevLett.109.121101} {\bibfield  {journal} {\bibinfo
  {journal} {Physical Review Letters}\ }\textbf {\bibinfo {volume} {109}},\
  \bibinfo {eid} {121101} (\bibinfo {year} {2012})},\ \Eprint
  {http://arxiv.org/abs/1205.4350} {arXiv:1205.4350 [astro-ph.HE]} \BibitemShut
  {NoStop}%
\bibitem [{\citenamefont {{Penrose}}(1969)}]{Penrose69}%
  \BibitemOpen
  \bibfield  {author} {\bibinfo {author} {\bibfnamefont {R.}~\bibnamefont
  {{Penrose}}},\ }\href@noop {} {\bibfield  {journal} {\bibinfo  {journal}
  {Nuovo Cimento Rivista Serie}\ }\textbf {\bibinfo {volume} {1}},\ \bibinfo
  {pages} {252} (\bibinfo {year} {1969})}\BibitemShut {NoStop}%
\bibitem [{\citenamefont {{Bardeen}}\ \emph {et~al.}(1972)\citenamefont
  {{Bardeen}}, \citenamefont {{Press}},\ and\ \citenamefont
  {{Teukolsky}}}]{Bardeen+72}%
  \BibitemOpen
  \bibfield  {author} {\bibinfo {author} {\bibfnamefont {J.~M.}\ \bibnamefont
  {{Bardeen}}}, \bibinfo {author} {\bibfnamefont {W.~H.}\ \bibnamefont
  {{Press}}}, \ and\ \bibinfo {author} {\bibfnamefont {S.~A.}\ \bibnamefont
  {{Teukolsky}}},\ }\href {\doibase 10.1086/151796} {\bibfield  {journal}
  {\bibinfo  {journal} {\apj}\ }\textbf {\bibinfo {volume} {178}},\ \bibinfo
  {pages} {347} (\bibinfo {year} {1972})}\BibitemShut {NoStop}%
\bibitem [{\citenamefont {{Wald}}(1974)}]{Wald74}%
  \BibitemOpen
  \bibfield  {author} {\bibinfo {author} {\bibfnamefont {R.~M.}\ \bibnamefont
  {{Wald}}},\ }\href {\doibase 10.1086/152959} {\bibfield  {journal} {\bibinfo
  {journal} {\apj}\ }\textbf {\bibinfo {volume} {191}},\ \bibinfo {pages} {231}
  (\bibinfo {year} {1974})}\BibitemShut {NoStop}%
\bibitem [{\citenamefont {{Kovetz}}\ and\ \citenamefont
  {{Piran}}(1975)}]{KovetzPiran75}%
  \BibitemOpen
  \bibfield  {author} {\bibinfo {author} {\bibfnamefont {A.}~\bibnamefont
  {{Kovetz}}}\ and\ \bibinfo {author} {\bibfnamefont {T.}~\bibnamefont
  {{Piran}}},\ }\href@noop {} {\bibfield  {journal} {\bibinfo  {journal} {Nuovo
  Cimento Lettere}\ }\textbf {\bibinfo {volume} {12}},\ \bibinfo {pages} {39}
  (\bibinfo {year} {1975})}\BibitemShut {NoStop}%
\bibitem [{Note1()}]{Note1}%
  \BibitemOpen
  \bibinfo {note} {There was a suggestion that an advanced civilization would
  power itself by dumping its garbage on a Kerr black hole and extracting, via
  the Penrose process, the black hole's rotational energy}\BibitemShut
  {NoStop}%
\bibitem [{\citenamefont {{Schnittman}}(2014)}]{Schnittman14}%
  \BibitemOpen
  \bibfield  {author} {\bibinfo {author} {\bibfnamefont {J.~D.}\ \bibnamefont
  {{Schnittman}}},\ }\href@noop {} {\bibfield  {journal} {\bibinfo  {journal}
  {ArXiv e-prints}\ } (\bibinfo {year} {2014})},\ \Eprint
  {http://arxiv.org/abs/1410.6446} {arXiv:1410.6446 [astro-ph.HE]} \BibitemShut
  {NoStop}%
\end{thebibliography}

%


\end{document}